\documentstyle[prb,aps,preprint]{revtex}
\begin{document}
\draft
\title{Adsorption of He atoms in external grooves of single wall
carbon nanotube bundles}

\author{Antonio \v{S}iber\thanks{E-mail: asiber@ifs.hr}}
\address{Institute of Physics, P.O. Box 304, 10001
Zagreb, Croatia \\
{\bf PUBLISHED IN PHYS. REV. B 66, 205406 (2002)}
}
\maketitle

\begin{abstract}
I calculate the quantum states for He atom in the potential of an external
groove of the single wall carbon nanotube bundle. The calculated ground
state energy is found to be in fair agreement with the experimental
estimate which suggests that the outer groove site is a preferential site
for the adsorption of He gas in the samples studied experimentally. I also
calculate the specific heat of
low-density $^4$He atom gas adsorbed in groove positions. The specific
geometry of the groove and its influence on the adsorbate quantum states
and specific heat are discussed.
\end{abstract}

\pacs{PACS numbers: 65.80.+n, 68.43.De, 68.65.-k }

\begin{center}
1. INTRODUCTION \\
\end{center}

Recent thermal desorption \cite{Teizer,Teizer2,Hallock} and specific heat
\cite{JHone,Las1} experiments on nanotube samples have detected $^4$He
atoms adsorbed in these samples. These experiments have attracted
attention due to the potential application of the carbon nanotube
materials for an efficient storage of gases
\cite{ColeCol,JLow,uptake}. In this respect, it is important to know the
details of the adsorption process and the preferential adsorption sites.
It is still not clear whether He
atoms adsorb in the interstitial channels between the close-packed
nanotubes within a bundle or in the external groove positions of the
bundle \cite{ColeCol,uptake} (see Fig.\ref{fig:fig1}). For
open-end tubes, He atoms could also adsorb within the tubes
\cite{cylind}. The
thermal desorption experiments reported in Ref. \onlinecite{Teizer} were
first successfully interpreted by assuming that He atoms occupy the
interstitial channels \cite{Teizer}, while later reevaluation of the
experimental data \cite{Teizer2} led authors of Ref. \onlinecite{Hallock}
to put forth the proposition that He atoms adsorb in the external grooves
of the nanotube bundle. Measurements of adsorption isotherms for
heavier gases (Xe, CH$_4$ and Ne)\cite{Migone} yielded that these gases do
not adsorb in the interstitial channels but probably in the external
grooves. By comparing their results for heavier gases with with the data
for $^4$He \cite{Teizer,Teizer2}, the authors of Ref. \onlinecite{Migone}
concluded that $^4$He atoms most probably adsorb also in the external
groove positions. The experimentally determined \cite{Teizer2} ground
state energy of $^4$He (-19.8 $\pm$ 1.5 meV) \cite{foot1} is significantly
higher than the ground state energy calculated for $^4$He atom adsorbed in
the nanotube interstitial channel (-27.8 meV in Ref. \onlinecite{Siber1},
-29.1 meV in Ref. \onlinecite{JLow} and -33.3 meV in
Ref. \onlinecite{uptake}). This
indeed suggests that He atoms may be adsorbed in the external groove
positions. The
theoretical estimate for
the ground state energy of $^4$He atoms adsorbed in the external groove
\cite{uptake} is -23.3 meV, still somewhat lower from the
experimental estimate. It is therefore of interest to examine the details
of adsorption of He atoms in the groove positions, in particular the
influence of very specific geometry of the groove position (see
Fig. \ref{fig:fig2}) on the bound state energies of He adsorbate.

The measurements of specific heat of He adsorbates in the 
nanotube samples could be of help to resolve the adsorption site
dilemma. In this respect, it should be of use to compare the theoretical
predictions for the specific heat with the experimental data. The
theoretical prediction
for the specific heat of low density (noninteracting) $^4$He gas adsorbed
in the interstitial channels has been presented in
Ref. \onlinecite{Siber1}. In this article I
shall consider the adsorption of He atoms in the external grooves of
nanotube bundle and calculate the specific heat for the low-density gas of
noninteracting He adsorbates.

The outline of the article is as follows. In Sec. 2, the
interaction potential confining the He adsorbate to the bundle external
groove is discussed. In Sec. 3, I shortly present a method of coupled
channels which is used to solve the single-particle Schr\"{o}dinger
equation in a potential of the groove. A short description of the
application of the method to the specific geometry of the groove is
given. The energies of quantum states of the adsorbate are calculated
together with the corresponding wave functions. In Sec. 4, the quantum
specific heat of $^4$He atoms
adsorbed in the groove positions is calculated. The results of the article
are summarized in Sec. 5.

\begin{center}
2. ADSORPTION POTENTIAL FOR He ATOM ON THE OUTER SURFACE OF THE BUNDLE \\
\end{center}

In Fig. \ref{fig:fig1}, a bundle containing 37 nanotubes is sketched. The
external groove and the interstitial channel adsorption
sites, together with the choice of the coordinate system are denoted. This
figure
suggests that the total He-bundle interaction potential, $V_t(x,y,z)$ can
be obtained as a sum of He-tube interactions, $V({\bf r}-{\bf r}_n)$,
\begin{equation}
V_t ({\bf r}) = \sum_{n} V({\bf r}-{\bf r}_n),
\label{eq:sumtube}
\end{equation}
where ${\bf r}=(x,y,z)$ and ${\bf r}_n$ are the radius vectors of He atom
and the
tube (labeled with index $n$), respectively.
The He-tube interaction
potential can in principle be obtained as a sum of effective He-C
interactions \cite{ColeCol,JLow,uptake,cylind,Siber1}. The thus obtained
He-tube potential would therefore exhibit corrugation along the tube
resulting from a discrete nature of the tube material. In what follows, I
shell neglect the corrugation of the He-tube potential. This
approximation effectively 
"smears" the carbon atoms to obtain a homogeneous, smooth tube surface. The
same approximation has been invoked in
Refs. \onlinecite{JLow,uptake,cylind}, but not in
Ref. \onlinecite{Siber1}. I shall later discuss the impact of this
assumption on final results.

It has been shown in
Ref. \onlinecite{cylind} that in the approximation of a smooth nanotube,
the He-single wall nanotube potential can be represented as
\begin{equation}
V(\rho) = 3 \pi \theta \epsilon \sigma^2 \left [ \frac{21}{32} \left(
\frac{\sigma}{R} \right )^{10} \eta ^{11} M_{11}(\eta) - \left(
\frac{\sigma}{R} \right )^{4} \eta ^{5} M_{5}(\eta) \right ],
\label{eq:tubepot}
\end{equation}
where $\epsilon$ and $\sigma$ are the energy and range parameter of the
effective He-C interaction assumed to be of a Lennard-Jones form. 
Variable $\rho$ is the distance of the He atom from the tube
axis, $\theta$ is the effective coverage of C atoms on a tube
surface (0.38 1/\AA$^2$), $R$ is the tube radius and
variable $\eta$ is defined as $\eta=R/\rho$ when $\rho>R$, which is a case
of interest to this work. Function $M_n(\eta)$ is defined as
\begin{equation}
M_n(\eta) = \int_{0}^{\pi} \frac{d \phi}{ (1+\eta ^2-2
\eta \cos\phi) ^{n/2}}.
\end{equation}
If the bundle is very large, the potential for He atom on the surface
of the bundle is almost periodic in $x$-direction (Fig. \ref{fig:fig1}),
due to the fact that there is a large number of tubes forming a
"flat" side of the bundle (the tubes denoted by thick line circles
in Fig. \ref{fig:fig1}). This suggests that we can write the total
He-bundle potential as a Fourier series,
\begin{equation}
V_t({\bf r}) = \sum_{G} V_{G}(z) \exp (i G x),
\label{eq:four1}
\end{equation}
where $G= 2 n \pi /a$, $n$ is integer, and $a$ is the distance between the
two neighboring tube axes as denoted in Fig. \ref{fig:fig1}. Note that the
total potential does not show dependence on $y$-coordinate due to the
fact that the tube discrete nature has been neglected.
By using Eq. (\ref{eq:sumtube}) in (\ref{eq:four1}), and neglecting
interactions of He atom with all tubes lying below the bundle side (which 
are vanishingly small due to a large diameter of carbon nanotubes), it is
straightforward to show that the Fourier components of the total
potential, $V_{G}(z)$, are given as
\begin{equation}
V_{G}(z) = \frac{2}{a} \int_{0}^{\infty} V(\sqrt{x^2+z^2}) \cos (Gx) dx.
\label{eq:fourcomp}
\end{equation}
The total potential obtained from Eqs. (\ref{eq:four1}) and
(\ref{eq:fourcomp}) exhibits complete two dimensional periodicity, i.e.,
it represents an infinitely long bundle side, as suggested by the dashed
circles in Fig. \ref{fig:fig1}. It is clear that the assumed periodicity
of the bundle side has no effect on the bound state energy at all, since
it is determined only by the two tubes surrounding a groove. Even for
higher bound states, the assumption of the potential periodicity does not
affect the eigen energies. This is again due to large diameter of
carbon nanotubes.

In what follows, I shall concentrate on (10,10) single wall nanotubes of
the so-called "armchair" type \cite{wrap}. The parameters of effective
Lennard-Jones He-C site potential I use are $\epsilon=1.45$ meV and
$\sigma=2.98$ \AA. These parameters were obtained from the so-called
combination rules\cite{Bruchbook} and were also used in
Ref. \onlinecite{uptake}. The He-tube potential
obtained from Eq. (\ref{eq:tubepot}) using the tube radius of $R=6.9$ \AA 
\hspace{0.7mm} has a well depth of -15.64 meV, and its minimum is
positioned at $\rho=$9.85 \AA. The total potential 
[Eq. (\ref{eq:sumtube})] is
plotted in Fig. \ref{fig:fig2} within the unit cell of the infinite
bundle side. The size of the unit cell is $a=2R+3.2$ \AA$=17$ \AA, in
agreement with experimentally determined tube-tube separation \cite{Thess}.

The expression for the He-tube potential in
Eq. (\ref{eq:tubepot}) is not convenient to calculate the Fourier
transforms in Eq. (\ref{eq:fourcomp}) in analytic fashion. In this
respect, I have found that the potential in Eq. (\ref{eq:tubepot}) can be
represented with excellent accuracy in the whole region of $\rho$
coordinate where the potential is smaller than $\sim$ 400 meV as
\begin{equation}
V(\rho) = A_1 \exp [-\beta_1 (\rho-\rho_0)] -  A_2 \exp [-\beta_2
(\rho-\rho_0)],
\label{eq:fitexp}
\end{equation}
with a set of parameters $A_1=3.4$ meV, $A_2$=18.81 meV, $\beta_1$=5.22
1/\AA, $\beta_2$=0.892 1/\AA, and $\rho_0$=9.85 \AA. Note that $\rho_0$ is
not
an independent parameter of this potential. It was introduced simply to
scale the values of $A_1$ and $A_2$
parameters to an acceptable range. In our choice of coordinate system,
$\rho=\sqrt{x^2+z^2}$. The functional form of the He-tube
potential in Eq. (\ref{eq:fitexp}) results in analytic expressions for
the Fourier components,
\begin{eqnarray}
V_{G}(z) &=& \frac{2}{a} \left [ A_1 \exp (\beta_1 r_0) \frac{\beta_1
z}{\sqrt{G^2+\beta_1^2}} K_1 (z \sqrt{G^2+\beta_1^2} ) \right .
\nonumber \\
&-& \left . A_2 \exp (\beta_2 r_0) \frac{\beta_2 
z}{\sqrt{G^2+\beta_2^2}} K_1 (z \sqrt{G^2+\beta_2^2} ) \right ],
\end{eqnarray}
where $K_1$ is the modified Bessel function of first order
\cite{Abram}. Note that the problem of He adsorption in external grooves
has
been effectively reduced to the problem of He adsorption on a very
corrugated surface made of carbon nanotubes. This enables us to use the
techniques developed to treat the problem of adsorption on surfaces
\cite{Bruchbook}.

\begin{center}
3. SOLUTION OF THE SCHR\"{O}DINGER EQUATION IN THE POTENTIAL OF THE
GROOVE \\
\end{center}

The single-particle Schr\"{o}dinger equation for a He atom in the groove
potential can be written as
\begin{equation}
-\frac{\hbar^2}{2M} \nabla ^2 \Psi({\bf r}) + V_t({\bf r}) \Psi({\bf r}) =
E \psi({\bf r}),
\label{eq:schrod}
\end{equation}
where $\Psi({\bf r})$ is the wave function of the adsorbate, $M$ and $E$ 
are its mass and energy, respectively. The symmetry of the potential
suggests that one can write the wave function in a form
\begin{equation}
\Psi({\bf r}) = \exp (iK_y y) \sum_G \xi_G (z) \exp [i(K_x+G)x],
\label{eq:razvoj}
\end{equation}
where the two-dimensional adsorbate wave vector in the $(x,y)$ plane is
denoted by ${\bf K}=(K_x,K_y)$. Substituting Eqs. (\ref{eq:razvoj}) and
(\ref{eq:four1}) in Eq. (\ref{eq:schrod}) one obtains a set of coupled
linear differential equations for functions $\xi_G (z)$,
\begin{equation}
\left ( \frac{d^2}{dz^2} + k_G^2 \right ) \xi_G (z) - \frac{2M}{\hbar^2}
\sum_{G'} V_{G-G'} (z) \xi_{G'} (z) = 0.
\label{eq:cceq}
\end{equation}
Here,
\begin{equation}
k_G^2 = \frac{2ME}{\hbar^2} - K_y^2 - (K_x+G)^2.
\end{equation}

The set of equations (\ref{eq:cceq}) typically occurs in scattering
problems \cite{Huts1,Johnson,SiberPhD,cceqref} where the equations (the
so-called coupled channel equations
\cite{cceqref}) are to be solved with a requirement
that $\xi_G (z)$ functions reduce to asymptotic scattering solutions in
the region where the interaction vanishes. In our case, we are interested
in states which vanish in the region where the potential has extremely
large values
and also in the region where the potential itself vanishes (bound
states). To solve these
equations for bound states, I adopt the procedure described in
Ref. \onlinecite{Huts1}. Basically, one numerically propagates the matrix
of $\xi_G (z)$ functions, (the log-derivative method of Johnson
\cite{Johnson} is used for the propagation), from the region of large
and small $z$ coordinate and requires that the matrix of solutions matches
smoothly at some predefined
coordinate $z_{fix}$. By systematically varying energy $E$ for fixed $K_x$, 
it is
possible to determine all the bound state energies of the
system \cite{Huts1}. Varying additionaly the value of $K_x$, one can
obtain the whole energy spectrum of the adsorbate. In a log-derivative
scheme for solving these
equations,
one actually does not propagate the wave functions ($\xi_G (z)$) and their
derivatives with respect to $z$ coordinate ($\xi_G' (z)$), but a
combination of type $\xi_G' (z)/\xi_G (z)$ \cite{cceqref,Huts1}, the
so-called log-derivative matrix. In this
way one avoids problems associated with the propagation of
solutions in classically highly forbidden regions \cite{cceqref,Huts1},
which is very important in
our case since the region of the groove is surrounded with highly
repulsive, classically forbidden regions (see Fig. \ref{fig:fig2}).
The details of this procedure for the geometry of 
interest here have been developed in Ref. \onlinecite{SiberPhD}. It
should be noted that the procedure used here to solve the Schr\"{o}dinger
equation is not the same as the one presented in Ref. \onlinecite{Siber1}
and used for a potential of the interstitial channel. In the case of the
interstitial channel potential, a much simpler scheme for the calculation
of bound states was used, due to the fact that the potential in that case
has much higher symmetry \cite{Siber1}.

The total
energy of the adsorbate can be written as \begin{equation}
E = \frac{\hbar^2 K_y^2}{2M} + E_m (K_x,K_y=0),
\end{equation}
so that it is sufficient to consider only $K_y=0$ case in solving the
equations (\ref{eq:cceq}). The index $m$ represents the adsorbate
"bands" obtained as solutions of Eqs. (\ref{eq:cceq}) with $K_y=0$.

It can be shown that the density of adsorbate bound states can be written
as
\begin{equation}
\rho_{2D}(E) = \sqrt{\frac{2M}{\hbar ^2}} \frac{2 L_x L_y}{(2 \pi)^2}
\sum_m
\int_0 ^{\pi/a} d K_x
\frac{\Theta[E-E_m(K_x,K_y=0)]}{\sqrt{E-E_m(K_x,K_y=0)}},
\label{eq:rho2d}
\end{equation}
where $L_x$ and $L_y$ are the dimensions of the normalization box in $x$
and $y$ directions, respectively, and $\Theta$ is the Heaviside step
function. Assuming that the solutions do not show appreciable dispersion
with $K_x$, i.e. that $E_m(K_x,K_y=0) \approx E_m(K_x=0,K_y=0) = E_m$, for
the density of states one can write
\begin{equation}
\rho_{2D}(E) = \sqrt{\frac{2M}{\hbar^2}} N_x \frac{L_y}{2 \pi} \sum_m
\frac{\Theta(E-E_m)}{\sqrt{E-E_m}},
\label{eq:rho1d}
\end{equation}
where $N_x$ is the number of tubes in the bundle side. Equation
(\ref{eq:rho1d}) is obviously a simple sum of effectively
one-dimensional densities of states for a single groove, reflecting a
situation where the adsorbate atom is strongly localized in a particular
groove region. It is therefore natural in this case to define a density of
states per unit length for a {\em single} groove as
\begin{equation}
g(E) = \frac{1}{L_y N_x} \rho_{2D}(E).
\label{eq:densscale}
\end{equation}

The potential in Fig. \ref{fig:fig2} cannot be accurately represented with
small number of Fourier components. This is due to the fact that the
groove region is very small on the scale of lattice constant, $a$. For
example, the equipotential contour of -30 meV has a mean radius of about
0.2 \AA. I have found that about 77 Fourier components with $G=\pm n 2
\pi /a$, $n=0,1,...,38$ are needed to represent the total potential with
an excellent accuracy in the whole region of importance.

In Fig. \ref{fig:fig3} a), the energy bands of a $^4$He
adsorbate are plotted. Note that the "bands" up until $\sim$ -11 meV are
indeed nearly dispersionless, so that
approximation for the density of states in Eq. (\ref{eq:rho1d}) holds with
great accuracy in that interval of energies. In Fig. \ref{fig:fig3} b) the
density of states obtained from Eq. (\ref{eq:densscale}) is plotted (the
details of band dispersion are included, i.e. Eq. (\ref{eq:densscale})
in combination with Eq. (\ref{eq:rho2d}) is used). As
expected from the geometry of the groove, the phase space is much larger
for higher He energies which is clear from the increase of the density of
states at
about -10 meV, an effect that is evident in Fig. \ref{fig:fig3} b). For
higher energies, the adsorbate becomes less confined to the groove region
and the energy levels become denser. The actual magnitude of the density
of states per unit length of a groove should be compared with the one
obtained in Ref. \onlinecite{Siber1} (Fig.2 of that reference) for He
adsorbed in an interstitial channel. From this comparison, it is evident
that the He atom is significantly less confined in a groove region than
in the interstitial channel region. This was of course expected prior to
any calculations. It is also visible in Fig. \ref{fig:fig3} that the
typical elements of the band structure calculations appear for bound
state energies higher than $\sim$ -11 meV, where the bands acquire
dispersion, which means that the probability of finding the adsorbate
in the region of a neighboring groove rises for these energies. It is also
interesting to note that the density of states becomes nearly constant
(as a function of energy) for energies between about -11 meV and -8 meV,
which is typical for the adsorbates nearly free in two dimensions
\cite{Bruchbook}. Thus, a transition from
the effectively one-dimensional adsorbate for low energies to effectively
two-and-more-dimensional adsorbate for higher energies, is evident in the
density of states.

Although the effects discussed in a previous paragraph can be traced
in the adsorbate density of states, it would certainly be of use to
visualize the
wave functions of the adsorbate. The log-derivative method can be used to
yield a precise information on the wave function \cite{Huts1}. In
Figs. \ref{fig:fig4} and \ref{fig:fig5}, we plot the probability densities
of the states denoted by letters from A to H in Fig. \ref{fig:fig3}. The
probability density, $P_m (x,z; K_x)$, for a
quantum state in a band $m$, characterised additionally with a wave
vector ${\bf K}=(K_x,K_y)$ is defined as
\begin{equation}
P_m (x,z; K_x) =  |\Psi_m (x,y,z; K_x)|^2,
\end{equation}
and the wave functions are normalized according to
\begin{equation}
\int_{0}^{a} dx \int_0^{\infty} dz |\Psi_m (x,y,z; K_x)|^2 = 1.
\end{equation}
Note that the probability density does not depend on $y$-coordinate and
$y$-component of the wave vector.

From Fig. \ref{fig:fig4} it can be seen that the higher energy states
become more delocalized in $x$, but most notably in $z$-direction,
i.e. the probability of finding an adsorbate atom in a region of large
$z$-coordinate rises with bound state energy. The dashed lines in
Figs. \ref{fig:fig4} and
\ref{fig:fig5} denote the 100 meV equipotential contour of He-bundle
interaction (see Fig. \ref{fig:fig2}). It is interesting to note that for
states G and H (two lower panels of Fig. \ref{fig:fig5}), the probability
density function becomes significantly delocalized over the whole surface
of the bundle. This reflects a highly delocalized adsorbate which is not
strongly confined to the groove region.

Occupation of the positions on the bundle surface and away from the
restricted region of the groove as a function of gas pressure has been
predicted in Ref. \onlinecite{Gatica1} for Ar adsorbates. In that case, as
the gas pressure rises, the coverage of the adsorbed gas increases in
almost discrete steps and the bundle surface eventually becomes covered by
a striped monolayer film. Upon further increase in pressure, a bilayer
adsorbate film forms. It should be emphasized that the change of effective
adsorbate dimensionality observed in this article is a quantum
effect related to a change of behavior of the single particle wave
function as the adsorbate bound state energy increases. Thus, this effect
is most easily observed as the sample temperature increases and the
occupation of excited bound states becomes probable. Similar effect has
been predicted for quantum gases adsorbed inside single-wall carbon
nanotubes. \cite{cylind,Gatica2} In that case, thermal excitation of the
azimuthal modes \cite{cylind} causes transition from 1D to 2D
behavior of adsorbates.

The ground state energy of $^4$He atom in a groove is $E_{\Gamma}$ =
-22.7 meV, in fair agreement with an experimental estimate
\cite{Teizer2,Hallock} (-19.8 $\pm$ 1.5 meV), and significantly higher 
from the ground state energy of $^4$He atom in an interstitial channel
(-27.9 meV) \cite{Siber1}. It is of interest now to examine the effect of
the details of the potential on the lowest bound state energy. In this
respect, for the effective He-C site potential, we adopt a slightly
different set
of parameters, used in Ref. \onlinecite{Siber1} and obtained from the
analysis of data on helium atom scattering from graphite \cite{Garcia}
($\epsilon$=1.34 meV and $\sigma=2.75$ \AA). This set of parameters
results in the ground state energy of -17.1 meV, which is in somewhat
better agreement with the experimental estimate. Interestingly enough,
relatively small change in the He-C binary potential parameters results in
a large change of ground state energy. This is due to a pronounced
sensitivity of the shape and magnitude of the total potential in the
region where the ground state is localized (see panel A of
Fig. \ref{fig:fig4}) on the details of the binary potential.
It is reasonable to expect that a more
refined He-tube potential would yield a better agreement with the
experimental estimate. It is also possible that an inclusion of
corrugation in the potential along the $y$-direction would produce a
somewhat lower ground state energy, as has been observed for He adsorption
on graphite surface \cite{CarCole}.

\begin{center}
4. SPECIFIC HEAT OF $^4$He ATOMS ADSORBED IN THE EXTERNAL GROOVES \\
\end{center}

As shown in Ref. \onlinecite{Siber1}, the isosteric specific heat $C/N$ of
low-density, noninteracting adsorbate gas, significantly delocalized in
one dimension ($y$-coordinate in our case), can be obtained from
\begin{equation}
\frac{C}{N} = L_y \frac{k_B}{(k_B T)^2} \frac{I_2 - I_1^2/I_0}{N},
\label{eq:specheat}
\end{equation}
where $N$ is the number of the adsorbate atoms within a {\em single}
external 
groove, $T$ is the sample temperature, $k_B$ is the Boltzmann constant,
and the integral quantities $I_j$ are defined as
\begin{equation}
I_j = \int_{E_{\Gamma}}^{\infty} g(E) E^j \exp \left[ \frac{E-\mu
(T)}{k_BT}
\right ] f^2(E,T) dE, \ j=0,1,2.
\label{eq:is}
\end{equation}
Here $E_{\Gamma}$ is the lowest allowed adsorbate energy (point A in
Fig. \ref{fig:fig3}), $\mu (T)$ is the chemical potential of the system
obtained from the
requirement of the adsorbate number conservation \cite{Siber1,CarCole},
and $f(E,T)$ is the Bose-Einstein distribution function
\begin{equation}
f(E,T) = \frac{1}{\exp \left( \frac{E-\mu}{k_B T} \right) -1 }.
\end{equation}  
The same equations appropriate for adsorption on surfaces have been
derived in Ref. \onlinecite{CarCole}. Note that when 
Eq. (\ref{eq:densscale}) in combination with Eq. (\ref{eq:rho2d}) is used,
Eq. (\ref{eq:specheat}) correctly accounts for the delocalization of the
adsorbate in $x$-direction.

The present approach does
not treat He-He interactions. Our results are therefore not applicable to
the whole range of temperatures and He atom concentrations. For
temperatures not so low and concentrations not so high that the He-He
interactions dominate the specific heat, this approach should yield
accurate results. A virial expansion approach to treat the He-He
interactions for He atoms adsorbed on graphite surface has been described
in Ref. \onlinecite{Siddon}. 

Fig. \ref{fig:fig6} shows the results for the isosteric specific heat
of the adsorbate gas as a function of sample temperature and for three
different linear densities ($N/L_y$) of adsorbates along the groove. All
three curves in Fig. \ref{fig:fig4} are very similar to each other but
very different from the corresponding curves when $^4$He atoms are assumed
to occupy interstitial channel positions (Figs. 3 and 4 of
Ref. \onlinecite{Siber1}). The difference is a consequence of the fact
that the He adsorbate in a groove becomes effectively two-dimensional for
lower sample temperature due to the fact that it is less efficiently
confined than in the case of adsorption in interstitial channel.
The observed difference also suggests that the measurements
of the specific heat of the low density He gas adsorbed in nanotube
samples could be used to determine the relevant He adsorption sites. The
results for the specific heat are expected to be less sensitive to the
corrugation of the potential than in the case of the interstitial channel
adsorption \cite{Siber1}. This is simply due to the fact that the
adsorbate is not completely surrounded by nanotubes. Therefore, the
adsorbate wave functions "sample" the corrugation in a smaller region of
space, particularly for higher energy states which are more extended in
$z$-direction. This is very different from the corresponding effect when
the adsorbate is localized in the interstitial channel region. In that
case, the higher energy states are more strongly influenced by the
corrugation due to the fact that their delocalization in the region of the
channel cross-section induces a larger overlap of the wave function with
the corrugated part of the confining potential.

\begin{center}
5. SUMMARY
\end{center}

Ground state energy of $^4$He atom adsorbed in the external groove
position
has been calculated and found to be in fair agreement with the
experimentally determined energy. All the excited states of
the adsorbate and the corresponding density of states have been calculated
from our approach. From these calculated quantities, we calculated the
specific heat of the adsorbate gas in the approximation of the
noninteracting adsorbates, applicable to low adsorbate densities. The
dependence of specific heat on temperature of the sample is markedly
different from the corresponding dependence when the adsorbates are
assumed to be in the interstitial channels \cite{Siber1}, which is due 
to less efficient confinement of the adsorbate in the groove region. The
measurements of the specific heat may thus provide an additional argument
in favor of the assignment of the external groove position as a
preferential
adsorption site for He atoms in the experiments performed in Refs. 
\onlinecite{Teizer,Teizer2,Hallock}.

\begin{figure}
\caption{
Geometry of the bundle of carbon nanotubes with designations of the
external
groove and interstitial channel adsorption sites and a choice of
coordinate system.
}
\label{fig:fig1}
\end{figure}

\begin{figure}
\caption{
The interaction potential (in meV) for He adsorbate in the
external groove region. The centers of the two tubes surrounding the
groove are positioned at $x=0.0$ \AA, $z=0.0$ \AA \hspace{0.7mm} and
$x=17.0$ \AA, $z=0.0$ \AA.
}
\label{fig:fig2}
\end{figure}

\begin{figure}
\caption{
a) Energy bands for $^4$He adsorbate in an external groove of
a bundle consisting of (10,10) nanotubes. The empty circles represent the
actually calculated points. The lowest energy band and the band connecting
G and H states have been drawn (full lines). b) The corresponding density
of states calculated from Eqs. (\ref{eq:densscale}) and (\ref{eq:rho2d}).
}
\label{fig:fig3}
\end{figure}

\begin{figure}
\caption{
The probability density functions of the states denoted by A,B,C and D in
Fig. \ref{fig:fig3}. The dashed lines represent the 100 meV equipotential
contour of the He-bundle potential.
}
\label{fig:fig4}
\end{figure}

\begin{figure}
\caption{
The probability density functions of the states denoted by E,F,G and H in
Fig. \ref{fig:fig3}. The dashed lines represent the 100 meV equipotential
contour of the He-bundle potential. Note the different scales of $x$ and
$z$ axes in upper and lower panels.
}
\label{fig:fig5}
\end{figure}

\begin{figure}
\caption{
Isosteric specific heat of a noninteracting gas of $^4$He atoms adsorbed
in external groove positions for three different adsorbate linear
densities. Dash-dotted line: $N/L_y=$0.033 1/\AA, Full
line: $N/L_y=$0.05 1/\AA, Dashed line $N/L_y=$0.1 1/\AA.
} 
\label{fig:fig6}
\end{figure}


\begin{references}
%
\bibitem{Teizer} W. Teizer, R.B. Hallock, E. Dujardin, and T.W. Ebbesen,
Phys. Rev. Lett. {\bf 82}, 5305 (1999)
%
\bibitem{Teizer2} W. Teizer, R.B. Hallock, E. Dujardin, and T.W. Ebbesen,
Phys. Rev. Lett. {\bf 84}, 1844(E) (2000)
%
\bibitem{Hallock} Y.H. Kahng, R.B. Hallock, E. Dujardin, and T.W. Ebbesen,
J. Low Temp. Phys. {\bf 126}, 223 (2002)
%
\bibitem{JHone} J. Hone, B. Batlogg, Z. Benes, A.T. Johnson, and
J.E. Fischer, Science {\bf 289}, 1730 (2000)
%
\bibitem{Las1} J.C. Lasjaunias, K. Biljakovi\'{c}, Z. Benes, J.E. Fischer,
and P. Monceau, Phys. Rev. B {\bf 65}, 113409 (2002)
%
\bibitem{ColeCol} M.M. Calbi, M.W. Cole, S.M. Gatica, M.J. Bojan, and
G. Stan, Rev. Mod. Phys. {\bf 73}, 857 (2001)
%
\bibitem{JLow} G. Stan, V.H. Crespi, M.W. Cole, and M. Boninsegni, J. Low
Temp. Phys. {\bf 113}, 447 (1998)
%
\bibitem{uptake} G. Stan, M.J. Bojan, S. Curtarolo, S.M. Gatica, and
M.W. Cole, Phys. Rev. B {\bf 62}, 2173 (2000)
%
\bibitem{cylind} G. Stan and M.W. Cole, Surf. Sci. {\bf 395}, 280 (1997)
%
\bibitem{Migone} S. Talapatra, A.Z. Zambano, S.E. Weber, and A.D. Migone,
Phys. Rev. Lett. {\bf 85}, 138 (2000)
%
\bibitem{foot1} The experimental uncertainty has been determined from
Fig. 2 of Ref. \onlinecite{Teizer2}.
%
\bibitem{Siber1} A. \v{S}iber and H. Buljan, Phys. Rev. B {\bf 66}, 075415
(2002)
%
\bibitem{wrap} N. Hamada, S.-I. Sawada, and A. Oshiyama,
Phys. Rev. Lett. {\bf 68}, 1579 (1992)
%
\bibitem{Thess} A. Thess, R. Lee, P. Nikolaev, H.J. Dai, P. Petit,
J. Robert, C.H. Xu, Y.H. Lee, S.G. Kim, A.G. Rinzler,
D.T. Colbert, G.E. Scuseria, D. Tomanek, J.E. Fischer, and R.E. Smalley,
Science {\bf 273}, 483 (1996)
%
\bibitem{Abram} M. Abramowitz and I.A. Stegun, {\em Handbook of
Mathematical Functions, With Formulas, Graphs, and Mathematical Tables} 
(Dover Publications, New York, 1972)
%
\bibitem{Bruchbook} L.W. Bruch, M.W. Cole, and E. Zaremba, {\em Physical
Adsorption: Forces and Phenomena} (Clarendon Press, Oxford, 1997)
%
\bibitem{Huts1} J.M. Hutson, Computer Physics Communications
{\bf 84}, 1 (1994)
%
\bibitem{Johnson} B.R. Johnson, J. Comput. Phys. {\bf 13}, 445 (1973)
%
\bibitem{SiberPhD} A. \v{S}iber, Ph.D. thesis, University of Zagreb,
2002. Downloadable at
\begin{verbatim}http://bobi.ifs.hr/~asiber/PhDThesis/\end{verbatim}
%
\bibitem{cceqref} M.H. Alexander, D.E. Manolopoulos, J. Chem. Phys. {\bf
86}, 2044 (1987); D.E. Manolopoulos, Ph.D. thesis, University of
Cambridge, 1988
%
\bibitem{Gatica1} S.M. Gatica, M.J. Bojan, G. Stan, and M.W. Cole,
J. Chem. Phys. {\bf 114}, 3765 (2001)
%
\bibitem{Gatica2} S.M. Gatica, G. Stan, M.M. Calbi, J.K. Johnson, and
M.W. Cole, J. Low. Temp. Phys. {\bf 120}, 337 (2000)
%
\bibitem{Garcia} N. Garcia, W.E. Carlos, M.W. Cole, and V. Celli,
Phys. Rev. B {\bf 21}, 1636 (1980)
%
\bibitem{CarCole} W.E. Carlos and M.W. Cole, Phys. Rev. B {\bf 21}, 3713
(1980)
%
\bibitem{Siddon} R.L. Siddon and M. Schick, Phys. Rev. A {\bf 9}, 907
(1974)
%
\end{references}
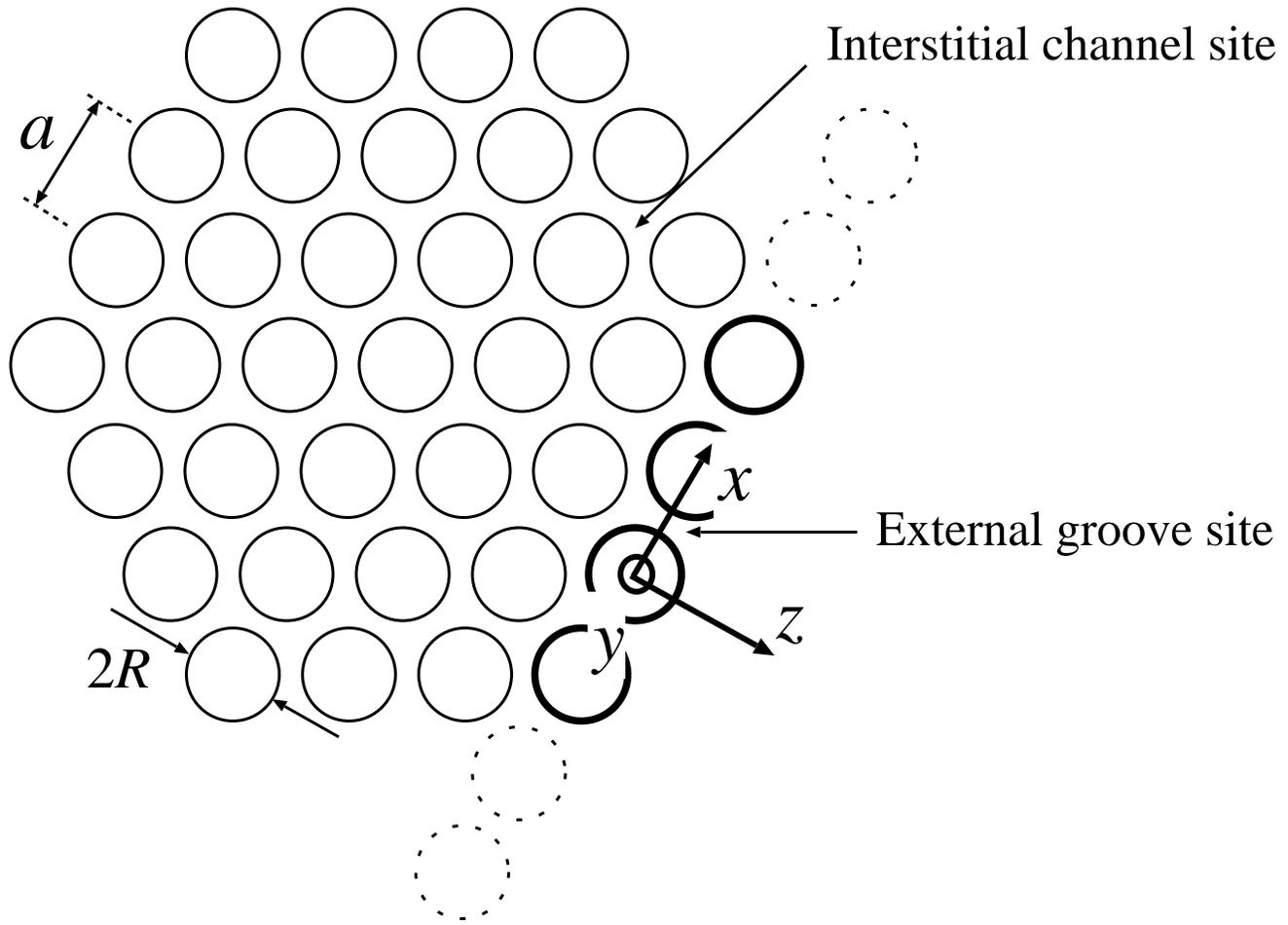
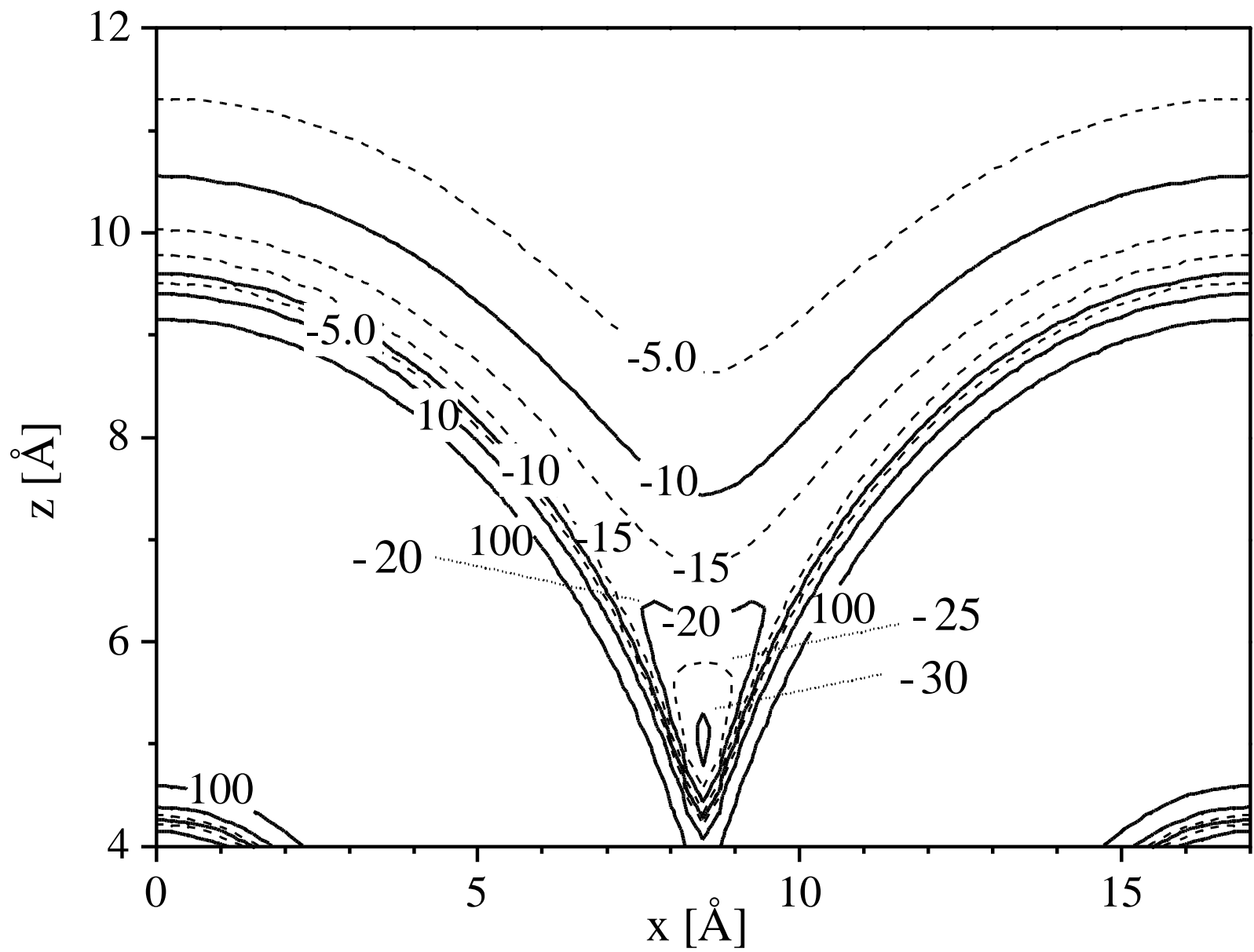
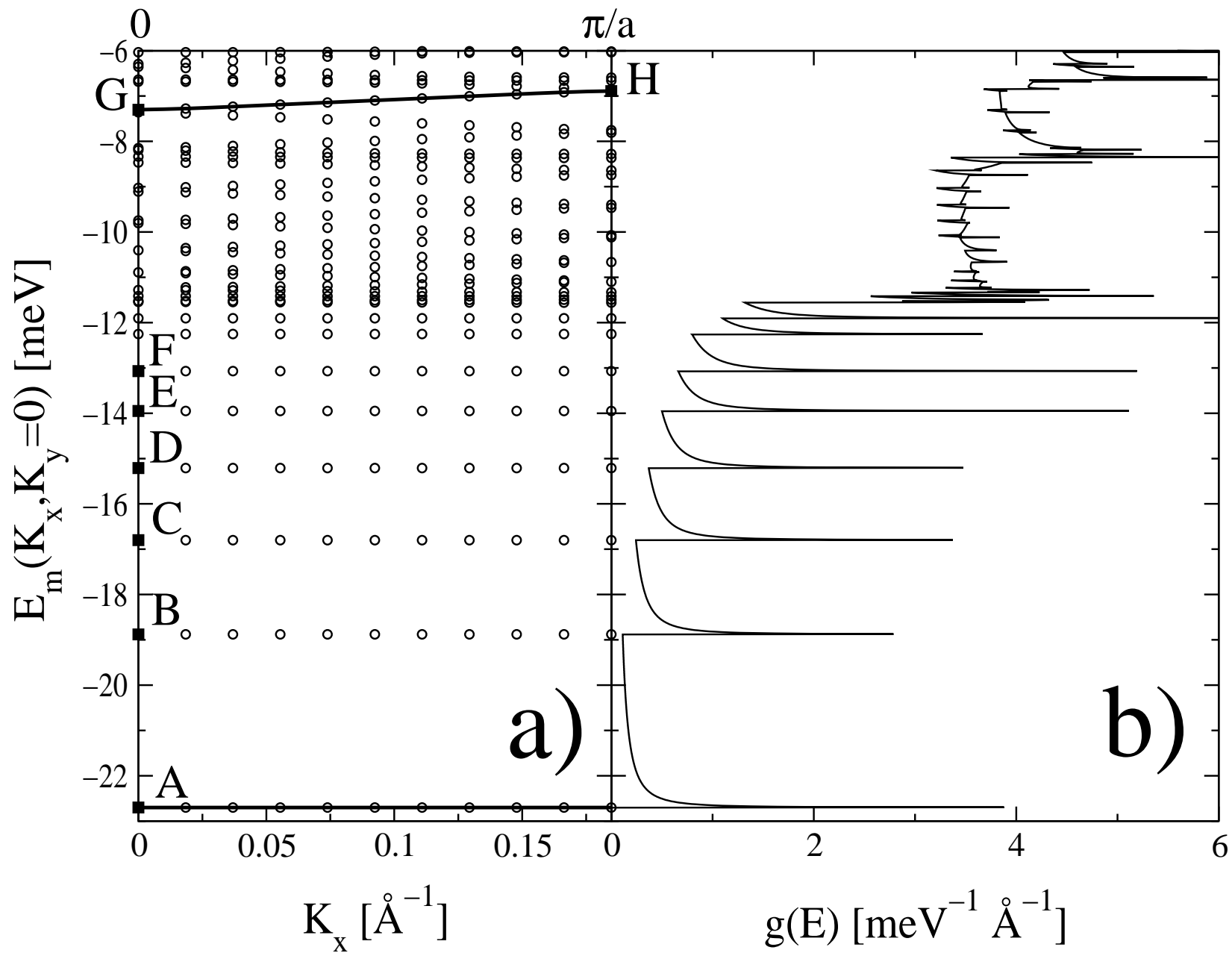
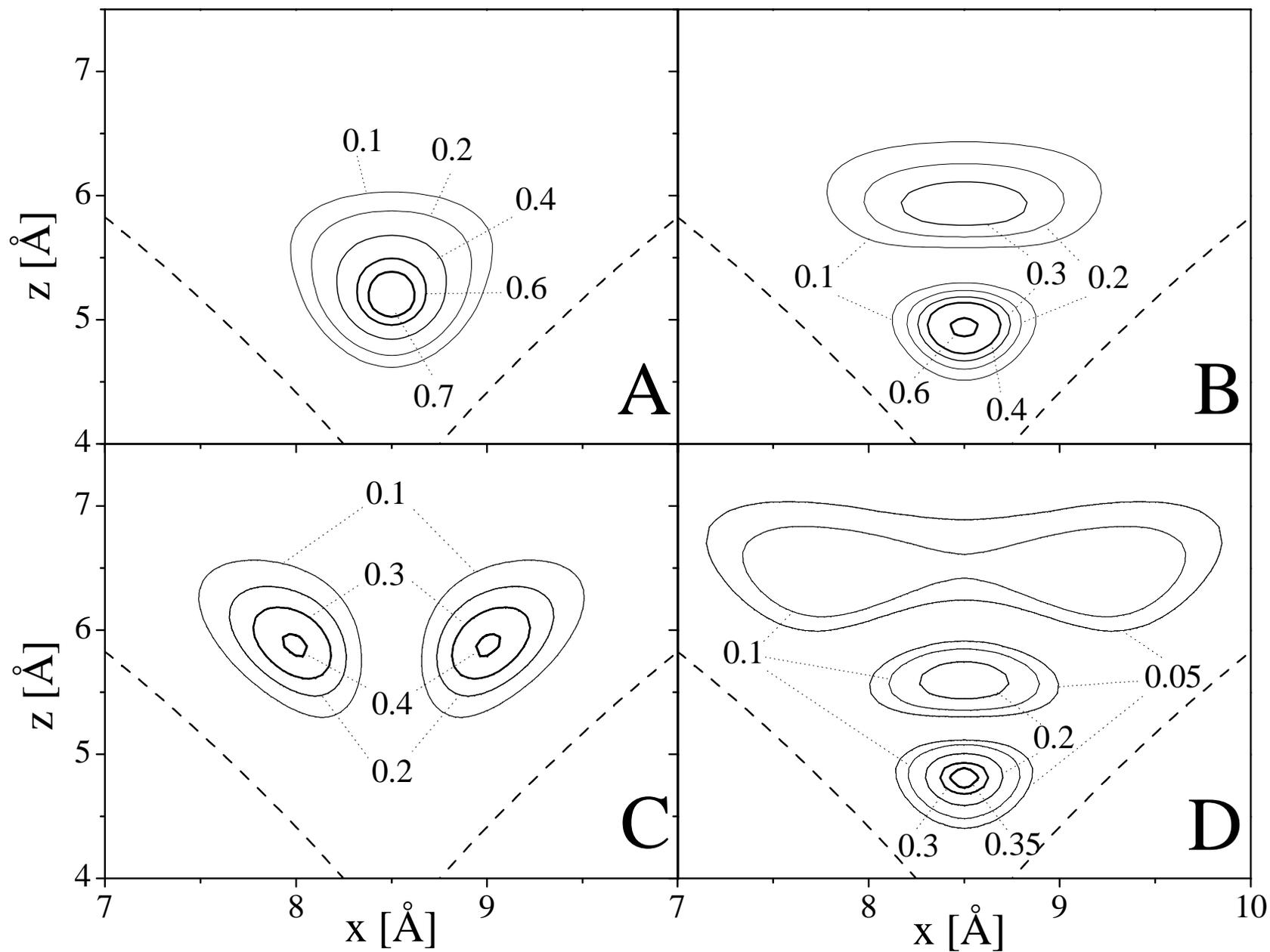
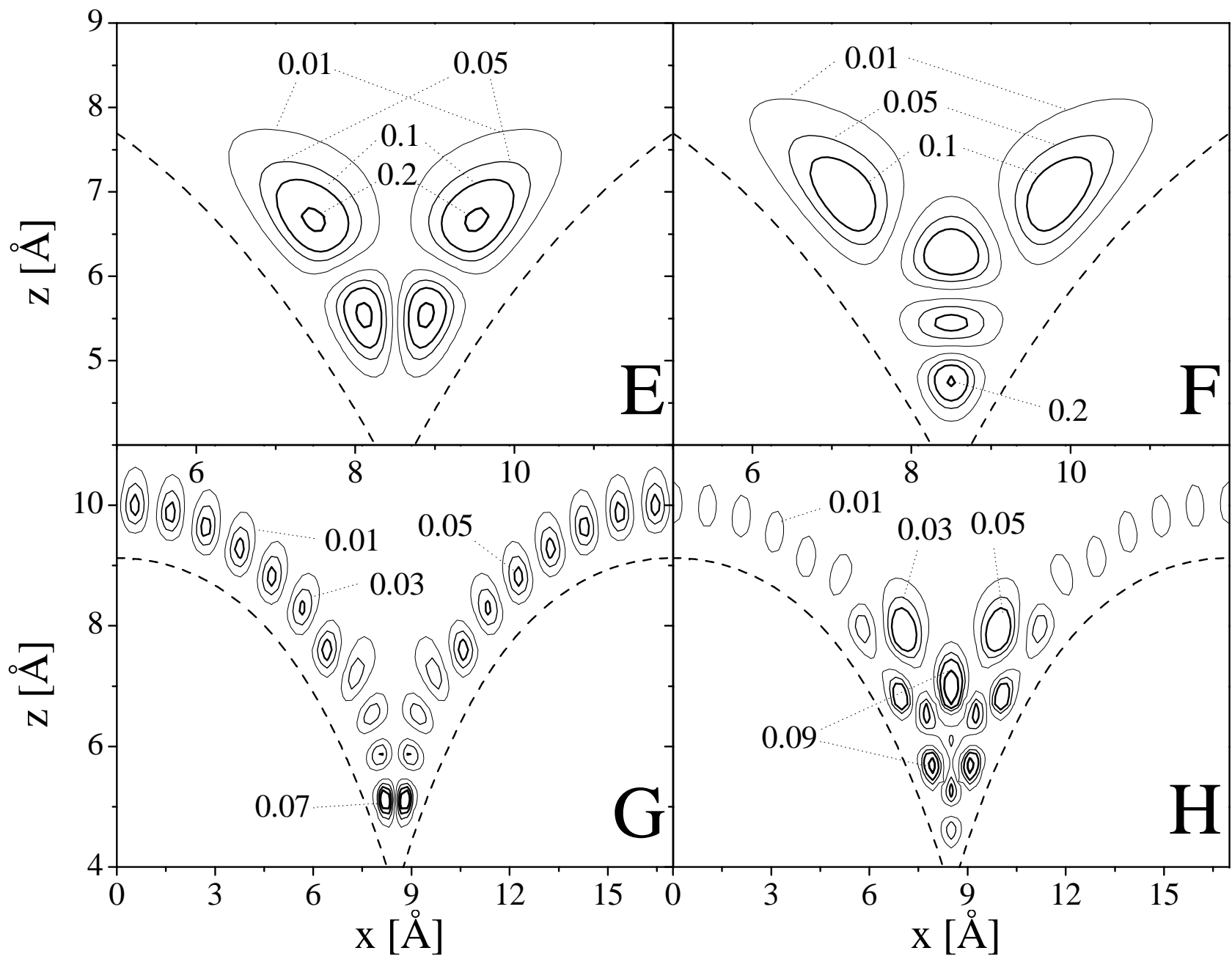
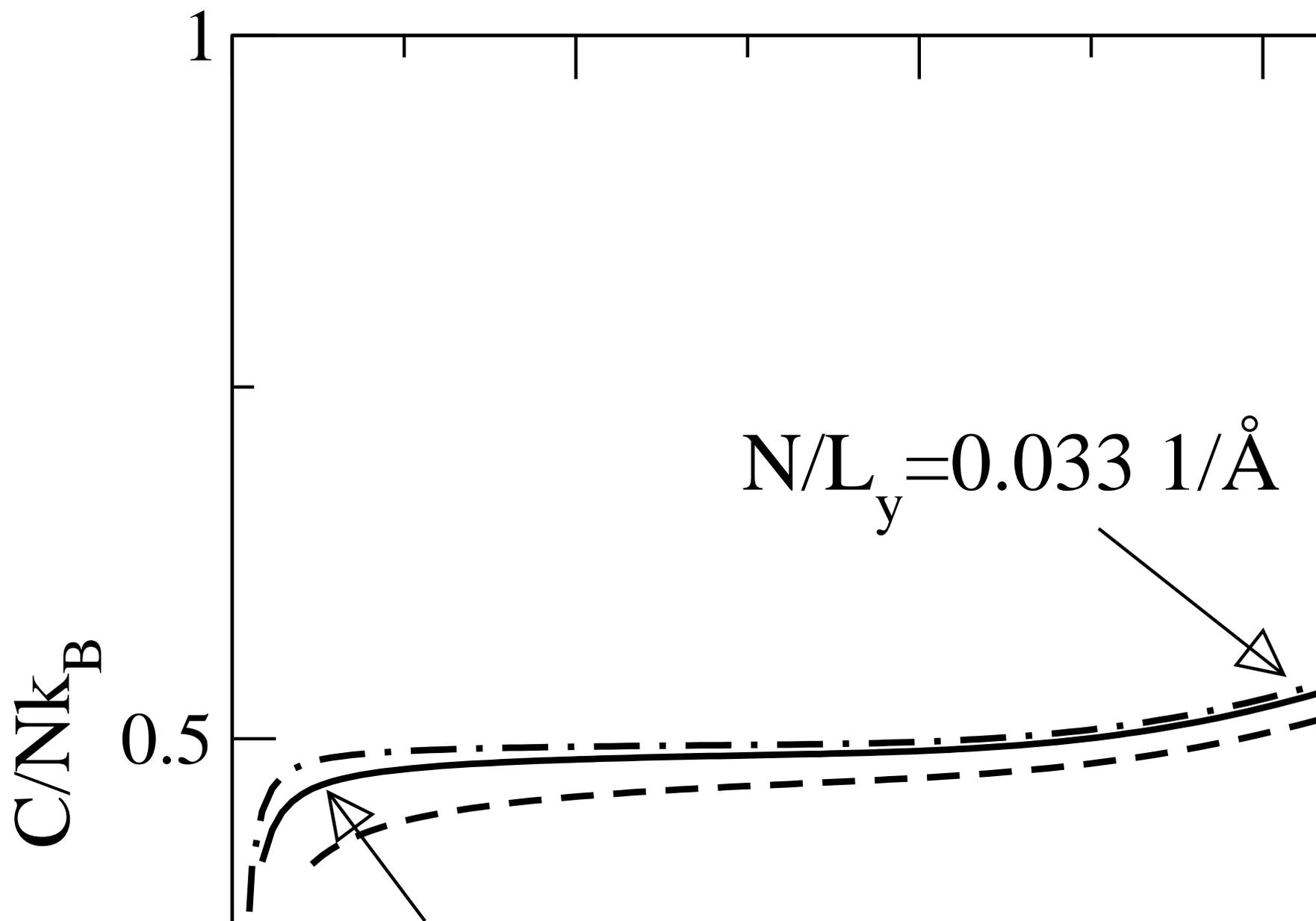
\end{document}